# Adsorption of ammonia at GaN(0001) surface in the mixed ammonia/hydrogen ambient - a summary of ab initio data

Paweł Kempisty[1,2], Stanisław Krukowski[1,3]

[1]Institute of High Pressure Physics, Polish Academy of Sciences, Sokołowska 29/37, 01-142 Warsaw, Poland

[3]Interdisciplinary Centre for Materials Modelling, Warsaw University, Pawińskiego 5a, 02-106 Warsaw, Poland

**Abstract**

Adsorption of ammonia at $NH_3/NH_2/H$ covered GaN(0001) surface was analyzed using results of ab initio calculations. The whole configuration space of partially $NH_3/NH_2/H$ covered GaN(0001) surface was divided into zones differently pinned Fermi level: at Ga broken bond state for dominantly bare surface (region I), at VBM for $NH_2$ and H covered (region II), and at CBM for $NH_3$ covered surface (region III). The extensive ab intio calculations show validity of electron counting rule (ECR) for all mixed coverage, for bordering these three regions. The adsorption was analyzed using newly identified dependence of the adsorption energy on the charge transfer at the surface. For region I and II ammonia adsorb dissociatively, disintegrating into H adatom and $HN_2$ radical for large fraction of vacant sites while for high coverage the ammonia adsorption is molecular. The dissociative adsorption energy strongly depends on the Fermi level at the surface (pinned) and in the bulk (unpinned) while the molecular adsorption energy is determined by bonding to surface only, in accordance to the recently published theory. The molecular adsorption is determined by the energy of covalent bonding to the surface. Ammonia adsorption in region III (Fermi level pinned at CBM) leads to unstable configuration both molecular and dissociative which is explained by the fact that Ga-broken bond sites are doubly occupied by electrons. The adsorbing ammonia brings 8 electrons to the surface, necessitating transfer of the electrons from Ga-broken bond state to Fermi level, energetically costly process. Adsorption of ammonia at H-covered site leads to creation of $NH_2$ radical at the surface and escape of $H_2$ molecule. The process energy is close to 0.12 eV, thus not large, but the inverse process is not possible due to escape of the hydrogen molecule.

# I. Introduction

Since the inception of MOVPE based nitride device technology, the physical properties of GaN(0001) surface and the growth of nitride layers have been investigated intensively, using both experimental and theoretical methods. These investigations led not only to the considerable understanding of the system but also to development of the new methods, which are now sufficiently mature to simulate the semiconductor (insulator's) surfaces fully accounting influence of the charged surface states and the fields close to the surface.[1-6] Recently, the dependence on the position of Fermi energy and doping was also accounted for.[6,7] These new methods led to discovery of new features fundamentally changing basic notions related to processes at semiconductor surfaces, introducing the dependence of adsorption energies on pinned and doping in the bulk for unpinned Fermi level surfaces.[6-8]

Basic properties of clean, hydrogen- and ammonia-covered GaN(0001) surface were obtained by DFT calculations without reference to doping or charged surface states.[9-14] According to these results the clean GaN(0001) surface does not undergo any reconstruction, the quantum surface states located in the bandgap are characterized by large dispersion of about 1.6 eV and are partially filled, i.e. the surface is metallic and the Fermi level is pinned.[10,11] More recent DFT simulations, using field-charge model indicate however that the clean GaN(0001) surface undergoes reconstruction to 2 x 1 row structure [4]. The energy difference between these two structures is very small and the numerous energy minima were found, that could lead to the relaxation termination in the local minimum, different from global minimum energy state. The states related to surface Ga atoms broken bonds are located at 0.6 eV below conduction band minimum (CBM), pinning Fermi level. Van de Walle and Neugebauer constructed phase diagram of GaN(0001) surface in ammonia ambient, using coordinates of chemical potential of hydrogen and nitrogen, finding regions of stability of several surface structures.[12,13] These results were confirmed by the DFT determination of the stable structures of polar and nonpolar GaN(0001) surfaces by Ito et al. [14]

The reaction of ammonia with the bare and H-covered surface was investigated by several groups by ab initio modelling,[15-20] Fritsch et al. simulated several configurations of GaN(0001) surface showing that ammonia adsorption is dissociative to H and $NH_2$ radical and transformation of GaN(0001) flat surface into p(2 x 2) vacancy reconstruction. The $NH_2$ radical was bound to gallium and H atom to gallium and nitrogen broken bonds respectively [15]. This reconstruction was not confirmed by any other calculations. The adsorption energy barrier was found to be of order of 0.5 eV. The DFT calculations by Pignedoli et al. also confirmed ammonia decomposition into H and $NH_2$ during adsorption at clean GaN(0001) surface.[15,17] In these calculations however, the surface remained flat with both products adsorbed on-top of Ga surface atoms. Similar, dissociative adsorption of ammonia at clean GaN(0001) surface, with the $NH_2$ radical adsorbed in the asymmetric bridge position, was obtained by Bermudez.[18,19] Kempisty et al. modeled adsorption of ammonia on highly H-covered GaN(0001) surface showing that ammonia is adsorbed molecularly in barrierless process [20]. The adsorption energy was high in excess of 2.5 eV. Recently intensive investigations of ammonia adsorption were conducted by the same authors [6]. They found that the ammonia adsorbed on relatively low H-occupied surface decomposes into $NH_2$ radical and H adatom in accordance with the results of Pignedoli et al. and Bermudez.[16-19] The adsorption energy is about 2.8 eV for the hydrogen coverage up to 0.75 ML, for higher coverage it is drastically reduced. Therefore in this region adsorption is molecular, locating the molecule in the 'on-top' position. The molecular adsorption energy, independent of the coverage, is close to 2.0 eV. For the adsorption on top of adsorbed H adatom, the Ga-H-N weak bond created, with the energy gain on adsorption of 0.8 eV, therefore reevaporation of the $NH_3$ molecule is possible. Finally the configuration dissociates into $NH_2$ radical and desorbing $H_2$ molecule.

The experimental investigations of the ammonia adsorption at GaN(0001) are not numerous. Supersonic molecular beams investigation of the ammonia adsorption indicated that ammonia adsorption is barrier-free process that proceeds via precursor-mediated mechanism, leading to ultimate dissociative stage in which the molecule disintegrates to $NH_x$ radicals [21]. It is not clear whether the surface was clean or hydrogen covered.

In the present work, the coverage diagram presenting Fermi level pinning at GaN(0001) surface will be constructed employing electron counting rule (ECR) [22] and DEF simulations. The ammonia adsorption using newly determined charge transfer dependence will be determined for entire configuration space of GaN(0001) in contact with $NH_3$-$N_2$-$H_2$ ambient.

## II. The simulation procedure

In the calculations reported below reported below, a freely accessible DFT code SIESTA, combining norm conserving pseudopotentials with the local basis functions, was employed.[23,24] The basis functions in SIESTA are numeric atomic orbitals, having finite size support which is determined by the user. The pseudopotentials for Ga, H and N atoms were generated, using ATOM program for all-electron calculations. SIESTA employs the norm-conserving Troullier-Martins pseudopotential, in the Kleinmann-Bylander factorized form [25]. Gallium 3d electrons were included in the valence electron set explicitly. The following atomic basis sets were used in GGA calculations: Ga (bulk) - 4s: DZ (double zeta), 4p: DZ, 3d: SZ (single zeta), 4d: SZ; Ga (surface)- 4s: TZ (triple zeta), 4p: TZ, 3d: SZ, 4d: SZ; N (bulk) - 2s: TZ, 2p: DZ;  N (surface)- 2s: TZ, 2p: TZ, 3d: SZ; H - 1s: QZ (quadruple zeta), 2p: SZ and H (termination atoms)  1s: DZ, 2p: DZ, 3d: SZ. The following values for the lattice constants of bulk GaN were obtained in GGA-WC calculations (as exchange-correlation functional Wu-Cohen (WC) modification [26] of Perdew, Burke and Ernzerhof (PBE) functional [27]: a = b = 3.2021 Å , c = 5.2124 Å. These values are in a good agreement with the experimental data for GaN: a = 3.189 Å and c = 5.185 Å [28] . All the presented dispersion relations are plotted as obtained from DFT calculations, burdened by a standard DFT error in the recovery of GaN bandgap. In the present parameterization, the effective bandgap for bulk GaN was 1.867 eV. In the case of the slab, the gap is additionally affected by localization in finite thickness increasing the gap to the following values: 20, 10 and 8 GaN layers: 1.925 eV, 2.123eV and 2.228 eV, respectively. Hence, in order to obtain a quantitative agreement with the experimentally measured values, all the calculated DFT energies that were obtained for 10 GaN layers slabs, should be rescaled by an approximate factor $\alpha = E_{g-exp}/E_{g-DFT}$=3.4eV/2.13eV ≈ 5/3 ≈ 1.6. Integrals in k-space were performed using a 3x3x1 Monkhorst-Pack grid for the slab with a lateral size 2x2 unit cell and only Γ-point for 4x4 slabs [29]. As a convergence criterion, terminating a SCF loop, the maximum difference between the output and the input of each element of the density matrix was employed being equal or smaller than $10^{-4}$. Relaxation of the atomic position is terminated when the forces acting on the atoms become smaller than 0.02 eV/Å.

Born-Oppenheimer approximation was used for determination of the energy barriers in which an effective procedure, based on nudged elastic band (NEB) method was applied.[30-32] The NEB method finds minimum energy pathways (MEP) between two stable points, which has to be predetermined first. The MEP is characterized by at least one first order saddle point, finding the energy barrier corresponding to activated energy complex approximation in chemical reaction kinetics. In the present formulation NEB module was linked to SIESTA package paving the way to fast determination of the energy and conformation of the species along the optimized pathways.

## III. The results

In the description below the coverage of GaN(0001) surface is calculated as the fraction of saturated Ga bonds by attached atoms. Thus zero coverage corresponds to the absence of species attached at the surface. Accordingly, the partial coverage is that in which a part of the surface sites has attached species while the rest have their broken bond nonsaturated. The full coverage is such that all Ga atoms have attached species, saturating all their bonds.

The following classification of the coverage is used: the presence of single species such as H, $NH_2$, etc is denoted as unary coverage. Consequently the two chemical species, such as H-$NH_2$, H-$NH_3$, $NH_2$-$NH_3$ will be denoted as binary coverage. And finally the ternary H-$NH_2$-$NH_3$ coverage is possible. In the all above cases either full or partial coverage could exist potentially, i.e. some part of Ga broken bonds could remain unsaturated.

As it was discovered recently [6,7], adsorption energy depends on the electric charge transfer between the solid bulk and the emerging states of the adsorbate. In the absence of the charge transfer, the adsorption energy attains the value, determined by the energy of the bonding. In addition, in some cases the adsorption energy could be possibly affected by the interactions with the neighbors. The notion that the adsorption energy may be affected by interaction with the neighbors on surfaces of metals and semiconductor was recognized and accepted for many years. The novel, recently introduced idea is the energy dependence on the electronic charge transfer [5,6]. The new phenomenon originates from the fact that the exchange of the electron between the Fermi level and the emerging surface state of different energies generates additional energy effect. These two states energy difference, may reach several electronvolts affecting adsorption energy considerably. This contribution changes the adsorption energy in case when the charge transfer is possible, i.e. when the empty states are available. The adsorption on semiconductor surfaces, involving charge transfer from/to the emerging state of the adsorbate depends on the Fermi level at the surface: pinned by surface state or free (unpinned) [6,7]. In the pinned case, the adsorption energy depends on the energy of the pinning surface state, which in most cases is a function of the chemical state of the surface, i.e. its coverage. In the case of the unpinned Fermi level surface, the adsorption energy depends on the position of the Fermi level in the bulk, i.e. the doping. Below we present the systematic discussion of the electronic properties of GaN surface and then the adsorption data are discussed.

**A. Uniform state of GaN(0001) surface**

The uniform state of the GaN(0001) is such that all sites of the surface are identical chemically i.e. the surface is either clean or fully covered by H, $NH_2$ and $NH_3$ species. The clean GaN(0001) surface was subject of intensive research. Therefore concise summary, including the features essential for the nonuniform coverage, is given here. In Figure 1 we present data on electronic properties of clean GaN(0001) surface. As it is shown, the Fermi level is pinned by Ga broken bond state, located in the bandgap, about 0.6 eV below conduction band minimum (CBM).

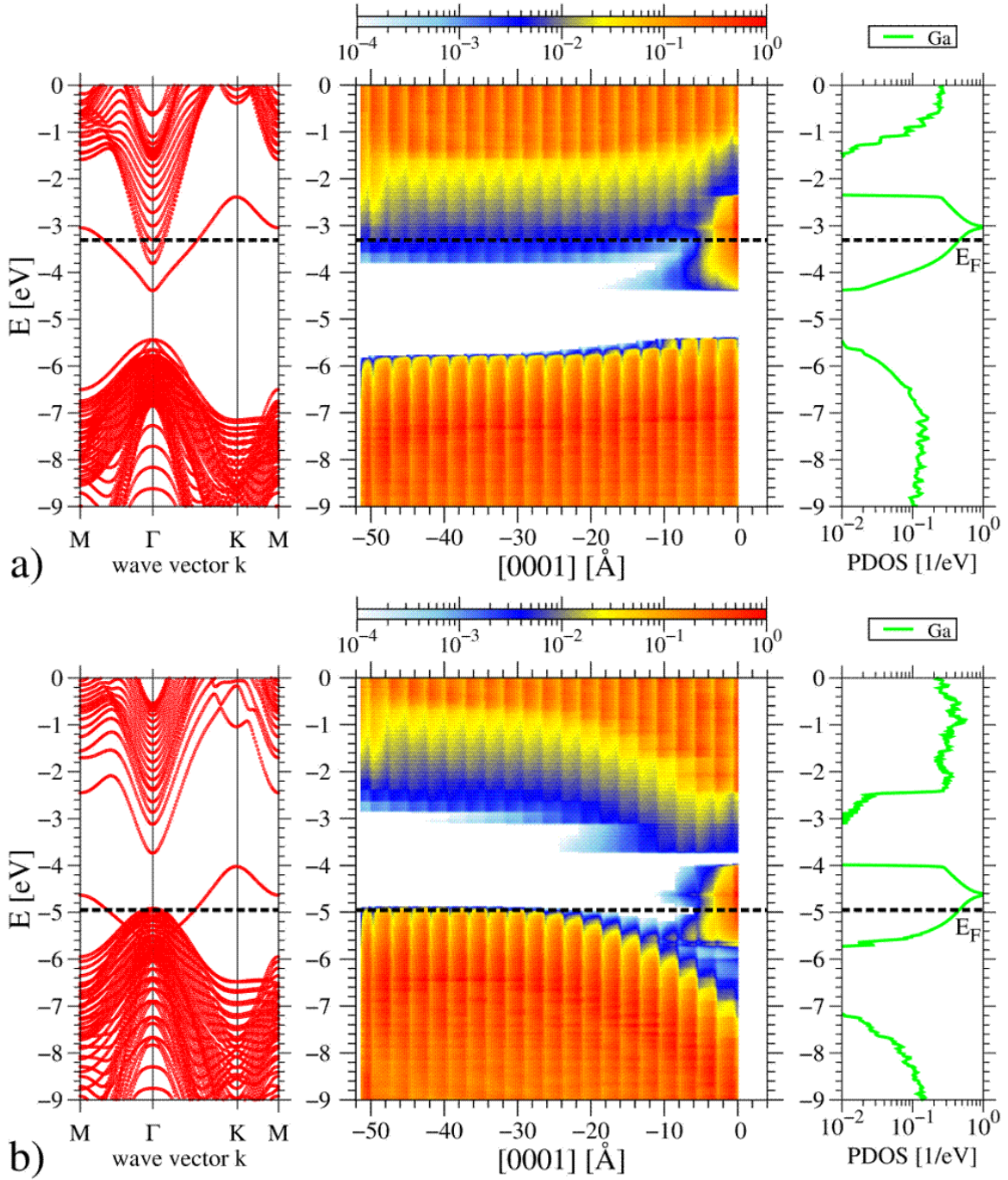

FIG. 1. Dispersion relations (left), space alignment of the bands derived from atom projected density of states (P-DOS) (middle left) and DOS projected on surface gallium atoms (middle right) and Crystal Orbital Hamilton Population (COHP [34]) of the N-H and Ga-N bonding (right), for clean GaN(0001) surface: a) n-type; b) p-type, obtained from 1 x 1 slab.

Accordingly, the type of the surface state (donor or acceptor) [5] depends on the doping in the bulk as shown in Figure 1. In the case of n-type doping, most frequently by silicon, the Fermi level in the bulk is at the donor defect level which is about 10 meV below CBM. Therefore the surface state is excessively charged becoming acceptor, with the band bend upwards by about 0.5 eV which leads to relatively small charge separation at the surface. In the case of p-type doping, the electronic charge is shifted to the bulk, resulting in strong excess positive charge on the surface state that behaves as a donor (Fig 1 b). The charge related downward band bending attains 3 eV, causing the immense surface charge effect. Naturally, such charge separation costs considerable energy. Therefore clean GaN surface is not likely to be encountered in the experiment. The simulation use relatively high temperature for electron distribution enabling screening by the band charge. Note also that the 1 x 1 slab was used, which does not allow emergence of 2 x 1 reconstruction [2,4].

The second uniform state considered is the GaN(0001) surface covered by hydrogen. As shown in Ref. 33, such coverage is extremely difficult to attain as it requires immense hydrogen pressures [33]. Nevertheless it was investigated theoretically, showing that hydrogen is attached in the atomic form in the position 'on-top' of Ga atoms, creating the state located in the valence band close to valence band maximum (VBM) as shown in Figure 2. Naturally, the Fermi level is pinned at the VBM, thus the band are almost flat for p-type and drastically bend up for n-type bulk. The possibility of existence of p-type material with hydrogen at the surface is doubtful as standard Mg acceptor creates a complex with hydrogen which is electrically neutral.

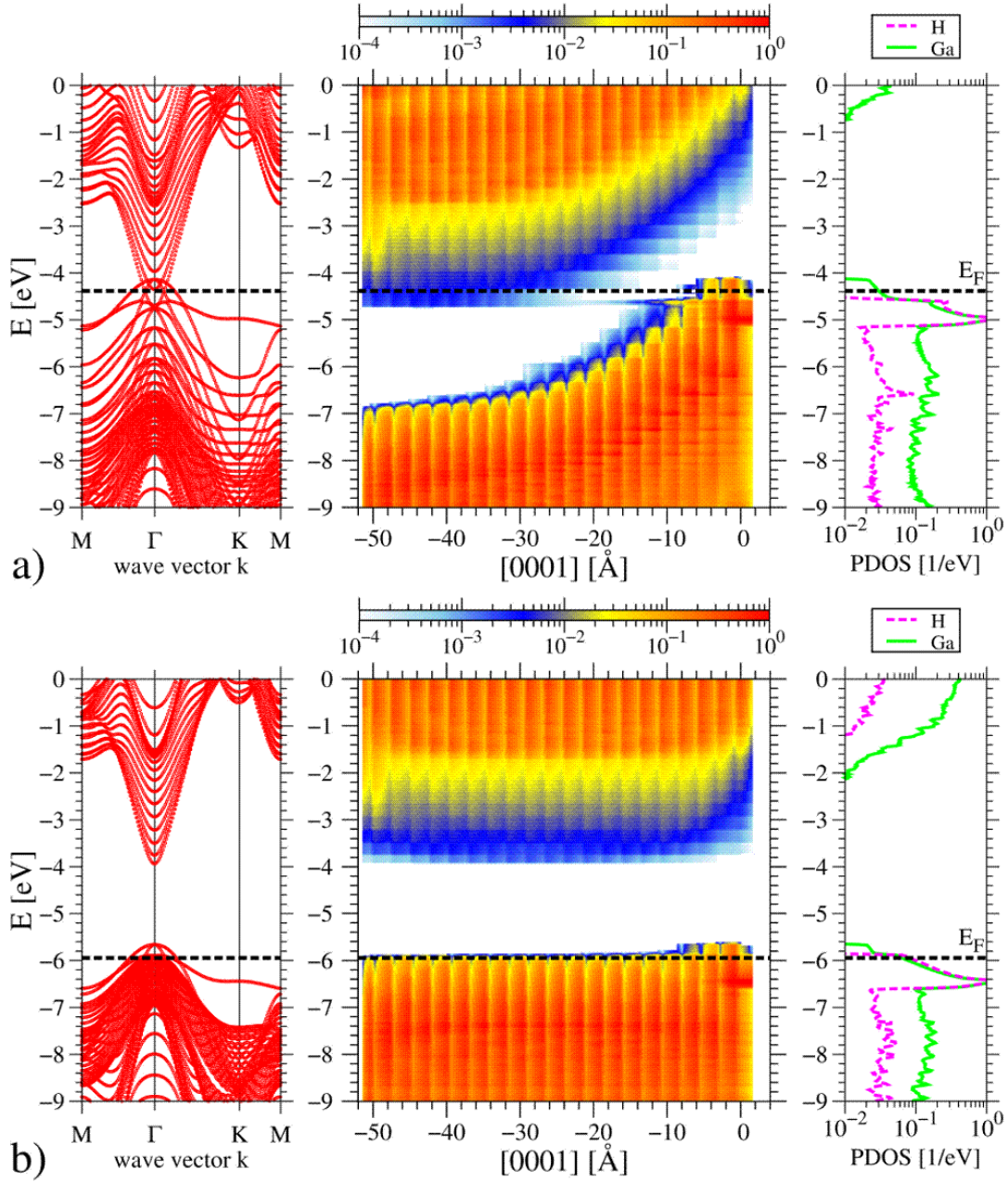

FIG. 2. Dispersion relations (left), space alignment of the bands derived from atom projected density of states (P-DOS) (middle left) and DOS projected on surface gallium and hydrogen atoms (middle right) and Crystal Orbital Hamilton Population (COHP [34]) of the N-H and Ga-N bonding (right), for fully hydrogen covered GaN(0001) surface: a) n-type; b) p-type, , obtained from 1 x 1 slab.

The third uniform case, of relevance to the present studies, is the GaN(0001) surface uniformly covered by $NH_2$ radicals. The electronic properties of such surface are presented in Figure 3. As it is shown, it is similar to the previous one as the Fermi level is pinned at VBM. As it is shown, this is related to the creation of the Ga-N bonding states located in the valence band (VB). The other molecular states of the radical, located deep in VB are completely occupied. In the result, the fully $NH_2$ covered surface is electronically identical to that covered by hydrogen: at p-type surface the band are essentially flat, at n-type they are bend upward over the entire gap, due to charged surface acceptor state.

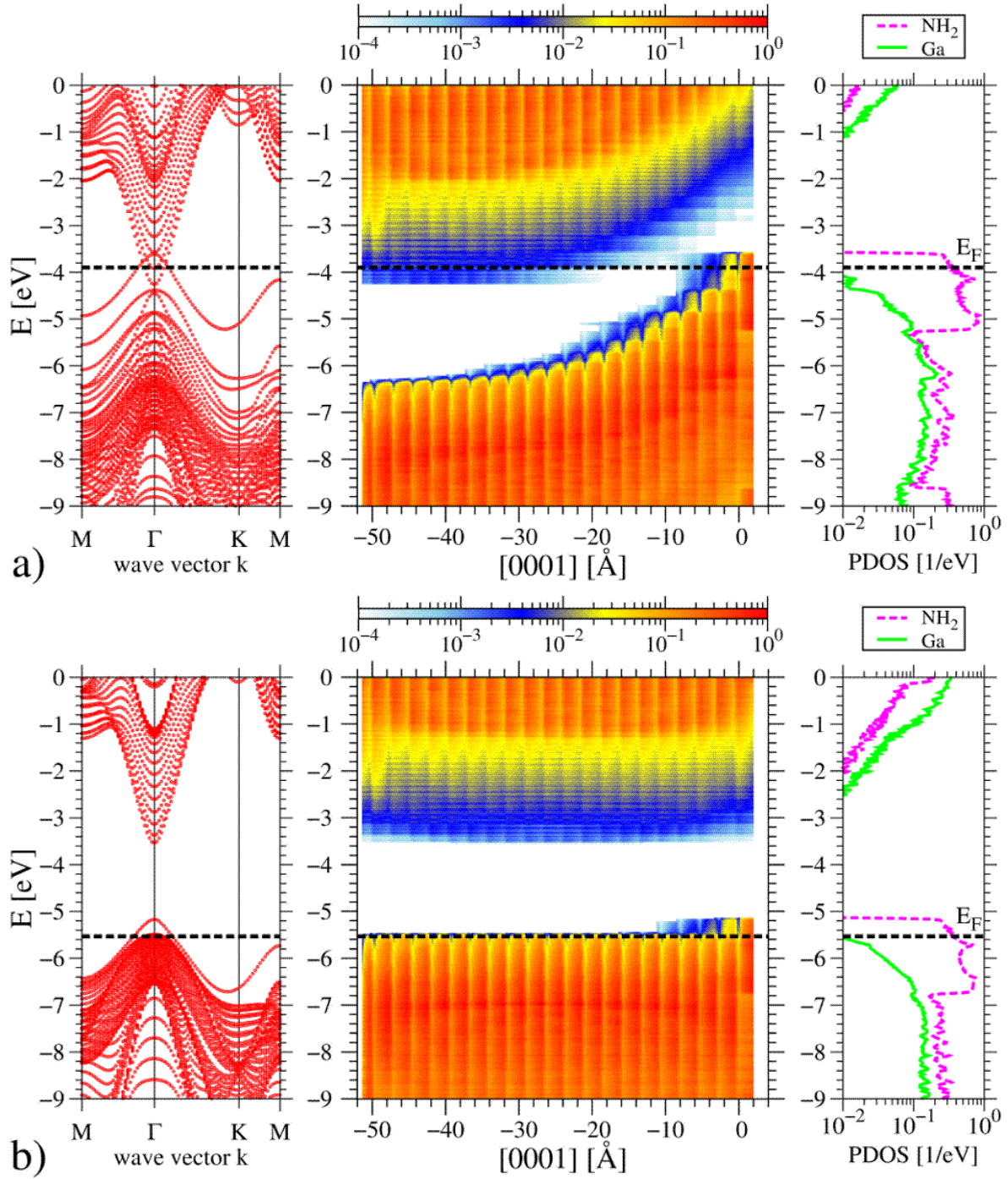

FIG. 3. Dispersion relations (left), space alignment of the bands derived from atom projected density of states (P-DOS) (middle left) and DOS projected on surface gallium and nitrogen-hydrogen (in $NH_2$ radicals) atoms (middle right) and Crystal Orbital Hamilton Population

(COHP [34]) of the N-H and Ga-N bonding (right), for fully $NH_2$-covered GaN(0001) surface: a) n-type; b) p-type, obtained from 1 x 1 slab.

As it is shown below, the surface fully covered by ammonia admolecules is radically different. The Fermi level is shifted up to energy deep in the conduction band (CB) so that the bands are almost flat for n-type gallium nitride (e.g. doped by Si). In the case of p-type material, the bands are severely bend downward, creating strong electric fields at the surface and increasing energy of the surface. The possibility of realization of such surface in the overwhelming presence of ammonia is doubtful.

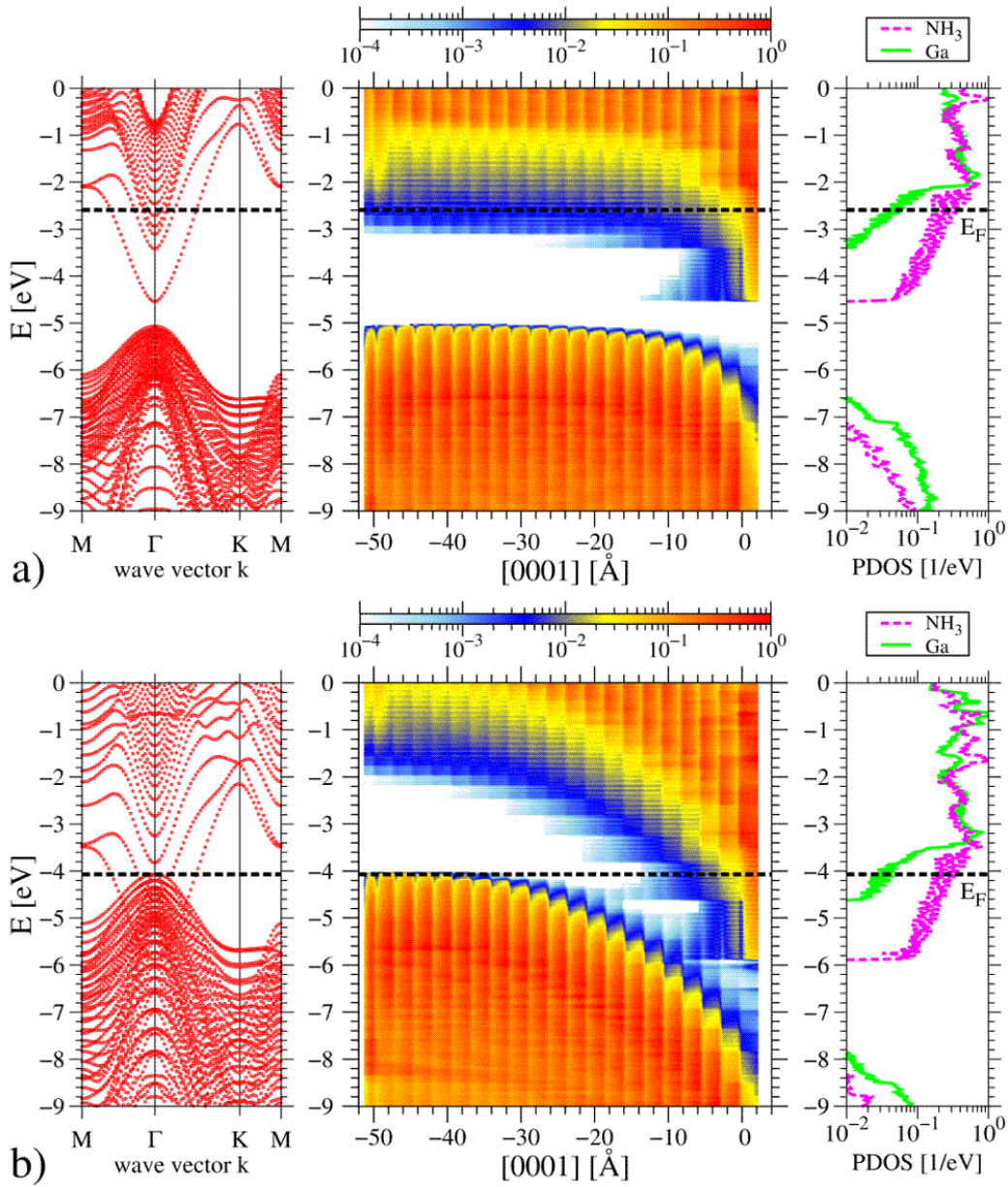

FIG. 4. Dispersion relations (left), space alignment of the bands derived from atom projected density of states (P-DOS) (middle left) and DOS projected on surface gallium and nitrogen-

hydrogen (in $NH_3$ admolecules) atoms (middle right) and Crystal Orbital Hamilton Population (COHP [34]) of the N-H and Ga-N bonding (right), for fully $NH_3$-covered GaN(0001) surface: a) n-type; b) p-type, obtained from 1 x 1 slab.

From the above discussion it follows that the Fermi level is pinned in all cases, in three different ways: at CBM, VBM or in the bandgap about 0.6 eV below CBM at Ga broken bond state. Therefore at some surface coverage's the transition of Fermi level occurs, i.e. it is pinned off and shifted from the one pinning surface state to the other one. The Fermi level pinning scenario indicates also that the only binary coverage without depinning is the binary H-$NH_2$ case for which the Fermi level is always pinned at VBM. Thus the coverage by H or $NH_2$ is equivalent electronically which simplifies the analysis as the coverage by H or by $NH_2$ will be treated as the same electronic state of the surface.

In order to determine the configurations of pinning absence, the electron counting rule (ECR) [22] is applied to determine the surface states and the conclusions are verified by DFT calculations in the next Section. Then, the diagram summarizing these changes is presented. Finally the configurations are used to determine the stable configurations of ammonia at the surface and the adsorption energies.

## B. Fermi depinning configurations at GaN(0001) surface - electron counting rule (ECR) state of the surface

The ECR analysis will not account the detailed analysis of bonding in GaN which is different from standard $sp^3$ bonding typical for III-V semiconductors [35,36]. In accordance to the ARPES results [35] and the DFT simulations [36] the valence band consists of the two subbands separated by the 5 eV wide gap. The upper subband is composed of *Ga 4s - 4p* and *N 2p* states [35]. Thus gallium is *$sp^3$* hybridized in bonding and nitrogen is not. The lower subband is composed of *Ga 3d* and *N 2s* states. The low energy states are filled first, thus the upper subband is the only one counted. For Ga surface, the built-in GaN crystal nitrogen atoms are not present at the surface and for gallium both ECR formulations are identical. Therefore we will follow standard Pashley arguments [22].

The clean GaN(0001) surface is terminated by the top Ga surface atom, which has missing N neighbor so that the single Ga bond has no overlap, i.e. gallium bond is broken. Due to that the energy of this state is located in the bandgap, 0.45 eV below CBM. According to ECR analysis 5/4 electron is missing, thus the Fermi level should be pinned by this state and that was confirmed by DFT data [4,7,10,11].

The unary coverage includes partial occupation by H adatoms, $NH_2$ radicals or $NH_3$ admolecules. The state associated with hydrogen on-top of Ga surface atom is located at VBM so it has to be occupied. Gallium contributes 3/4 electron and hydrogen 1 electron. Therefore the Ga broken bond site contributes 3/4 electron while the one with attached hydrogen 1 + 3/4. These electrons occupy two quantum states associated with hydrogen while Ga broken bonds are empty. Thus the charge redistribution balance gives

$$\frac{3}{4}\alpha + \left(1 + \frac{3}{4}\right)\beta = 2\beta + 0\alpha \qquad (1)$$

where α and β are fractions of donors (Ga broken bonds) and acceptors (Ga saturated by hydrogen) respectively. Naturally the fraction of both sites is normalized to unity, i.e.:

$$\alpha + \beta = 1 \qquad (2)$$

Effectively the H occupied site behaves as acceptor of 1/4 electron which has to be provided from Ga broken bond state in the bandgap, i.e.:

$$\frac{3}{4}\alpha = \frac{1}{4}\beta \qquad (3)$$

Thus the ratio of donors and acceptors is 1: 3 and accordingly ECR is attained for β = 0.75 ML hydrogen coverage, the fraction of empty sites is: α = 0.25.

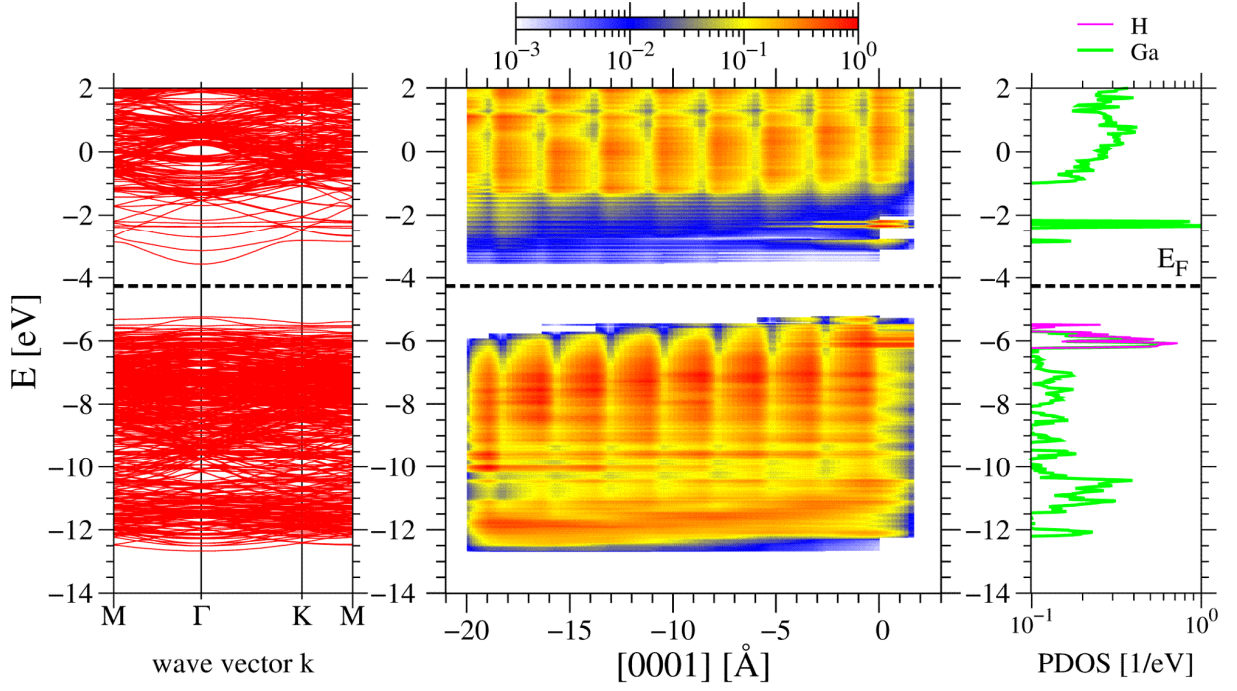

FIG. 5. Dispersion relations (left), space alignment of the bands derived from atom projected density of states (P-DOS) (middle) and DOS projected on surface gallium and nitrogen-hydrogen (in $NH_3$ admolecules) atoms (right) for 0.75 ML H-covered GaN(0001) surface.

The above ECR prediction is in accordance with the results of DFT simulations presented in Ref.4 and also in Fig5. Thus the critical unary coverage, corresponding to flat bands at the surface and the Fermi level not pinned at the surface is: 0.75 ML hydrogen coverage as shown in Fig 5.

Similar analysis may be made for the GaN(0001) surface, fractionally covered by $NH_2$ radicals. According to Figure 3, all $NH_2$ states (eight states) are degenerate with valence band thus they have to be occupied. Nitrogen provides five electrons and two other electrons come from the two hydrogen atoms giving in total 7, which has to be added to 3/4 electron from the surface Ga atom. These electrons have to fill eight states associated with $NH_2$ radical, Ga broken bond state is empty:

$$\frac{3}{4}\alpha + \left(7 + \frac{3}{4}\right)\beta = 8\beta + 0\alpha \qquad (4)$$

as before α and β are fractions of donors (Ga broken bonds) and acceptors (Ga with $NH_2$ radicals attached) respectively. As before, the sum of fractions of both sites is normalized to unity, fulfilling Eq. 2. As in the case of hydrogen coverage, the ECR state is attained for 0.75 ML NH$_2$ coverage ($\alpha = \frac{1}{4}$, $\beta = \frac{3}{4}$).

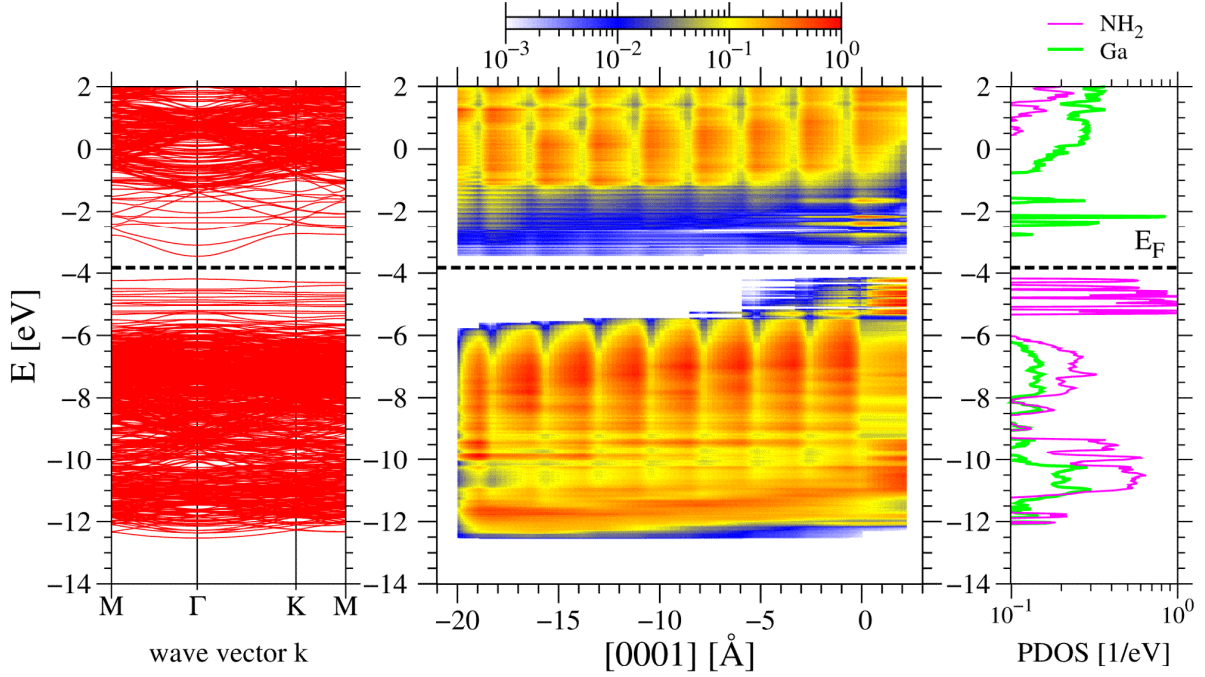

FIG. 6. Dispersion relations (left), space alignment of the bands derived from atom projected density of states (P-DOS) (middle) and DOS projected on surface gallium and nitrogen-hydrogen (in NH$_2$ radicals) atoms (right), for 0.75 ML NH$_2$-covered GaN(0001) surface.

As above, the DFT result proves that flat bands at the surface are attained for 0.75 ML coverage by NH$_2$ radicals, thus confirming full analogy between H and NH$_2$.

Finally, the partial unary coverage by NH$_3$ admolecules has to be considered in which the upper filled states are degenerate with conduction band while lower energy states are degenerate with the valence band (Figure 4). In the case of NH$_3$ admolecule, nitrogen atom contributes 5 electrons, three hydrogen atoms 3 which has to be added to the surface Ga atom broken bond contribution of 3/4 electron. These electrons have to occupy 8 NH$_3$ states and 2 Ga-H states:

$$\left(8 + \frac{3}{4}\right)\alpha + \frac{3}{4}\beta = 8\alpha + 2\beta \quad (5)$$

as before α and β are fractions of donors (Ga with $NH_3$ admolecules) and acceptors (Ga broken bonds) respectively. Total fraction of both sites is normalized to unity, following Eq. 2. Effectively 5/4 electron has to be taken from neighboring $NH_3$ admolecules, i.e. electron transfer equation gives:

$$\frac{3}{4}\alpha - \frac{5}{4}\beta = 0 \qquad (6)$$

Therefore the ratio of the $NH_3$ filled to empty sites is 5 : 3 so ECR is attained for α = 0.625 $NH_3$ ML coverage, the remaining β = 0.375 fraction is empty. The Fermi energy is located between Ga broken bond and CBM.

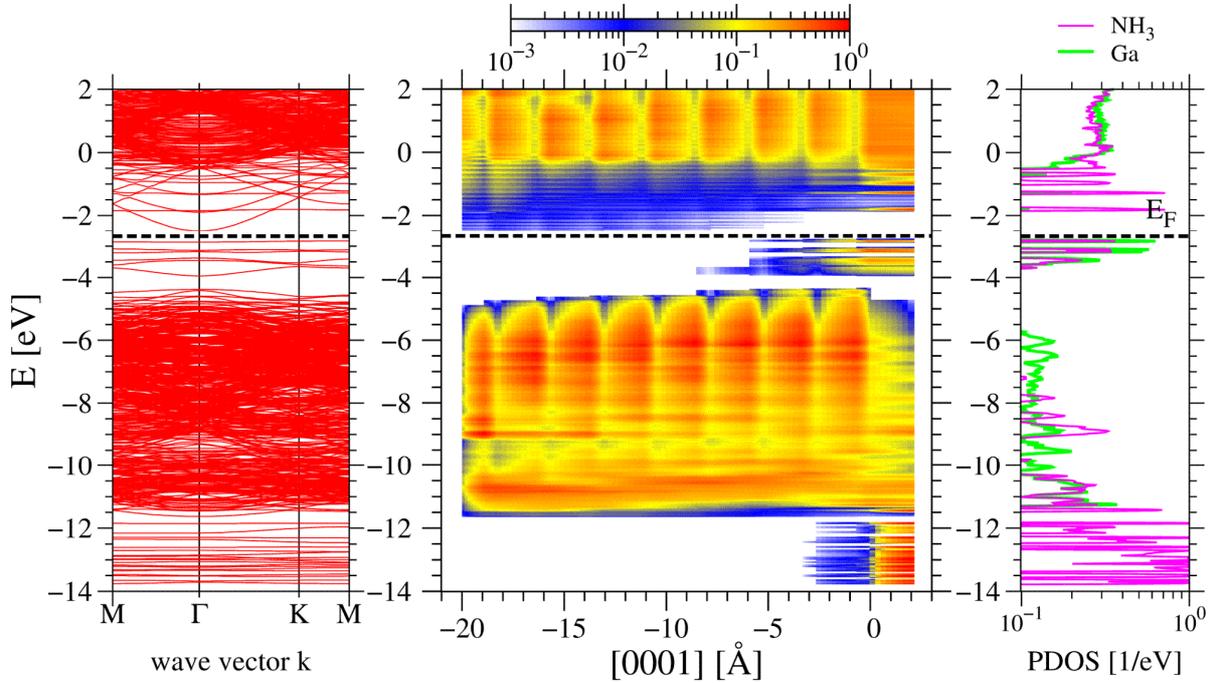

FIG. 7. Dispersion relations (left), space alignment of the bands derived from atom projected density of states (P-DOS) (middle) and DOS projected on surface gallium and nitrogen-hydrogen (in $NH_3$ admolecules) atoms (right) for 0.625 ML $NH_3$-covered GaN(0001) surface. The flat band termination surface is used.

As shown in Figure 7, the DFT results confirm the Fermi level position between CBM and Ga broken bond state, that is located 0.6 eV below CBM, i.e. in the narrow energy strip. Thus the obtained DFT results are in accordance with ECR analysis.

The full binary $NH_2$ - $NH_3$ coverage attains depinning Fermi level at approximately 0.25ML $NH_3$ content. This may be explained by ECR rule [22]. $NH_3$ admolecule contributes 5 electrons from nitrogen atom, 3 from three hydrogen atoms and 3/4 from Ga broken bond while $NH_2$ radical contributes 5 electrons from nitrogen atom, 2 from three hydrogen atoms and 3/4 from Ga broken bond, which has to occupy 8 states associated with $NH_3$ admolecule and similarly with $NH_2$ radical:

$$\left(8+\frac{3}{4}\right)\alpha + \left(7+\frac{3}{4}\right)\beta = 8\alpha + 8\beta \qquad (7)$$

α and β are fractions of donors (Ga with $NH_3$ admolecules) and acceptors (Ga with $NH_2$ radicals) respectively. Total fraction of both sites is normalized to unity, in accordance to Eq.2. The ECR state is attained for the following coverage: α = 0.25 $NH_3$ ML and β = 0.75 $NH_2$ ML. The electronic properties of GaN(0001) surface under such coverage, obtained from DFT simulations, are presented in Figure 8, confirming the above predictions.

Essentially the same result may be obtained for full binary $NH_3$ - H coverage, accounting that hydrogen occupied site provides only 1 electron from hydrogen and 3/4 from gallium, but accepts only 2 electrons:

$$\left(8+\frac{3}{4}\right)\alpha + \left(1+\frac{3}{4}\right)\beta = 8\alpha + 2\beta \qquad (8)$$

α and β are fractions of donors (Ga with $NH_3$ admolecules) and acceptors (Ga with H adatom) respectively. Total fraction of both sites is normalized to unity, in accordance to Eq. 2. Since, effectively the shortage of 1/4 electron is common for H adatom and $NH_2$ radical, the results is identical α = 0.25 $NH_3$ ML and β = 0.75 H ML confirming analogy between $NH_2$ and H coverage.

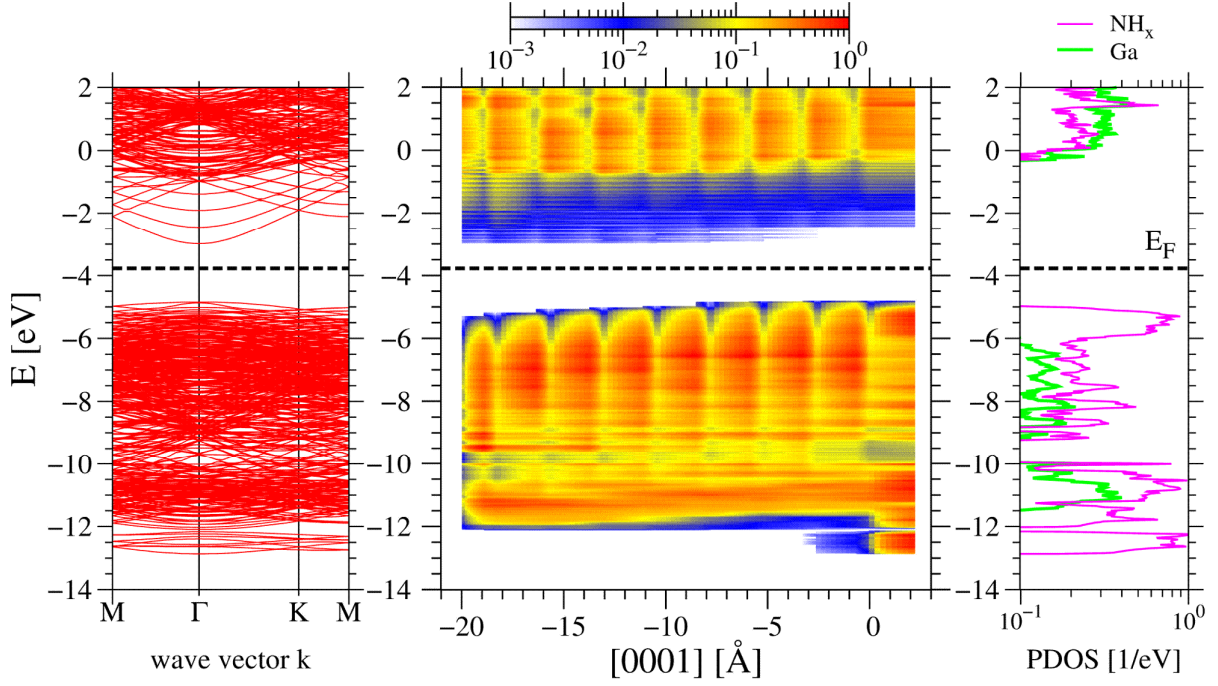

FIG. 8. Dispersion relations (left), space alignment of the bands derived from atom projected density of states (P-DOS) (middle) and DOS projected on surface gallium and nitrogen-hydrogen (in $NH_3$ admolecules) atoms (right) for 0.25 ML $NH_3$ - 0.75 ML $NH_2$ covered GaN(0001) surface.

Finally, partial binary $NH_2$ - $NH_3$, H - $NH_3$ and also ternary H - $NH_2$ - $NH_3$ coverage may be analyzed. In this case the three pinning states are possible: at VBM, CBM and Ga broken bond quantum state. The first ECR appears for the Fermi level positioned between VBM and Ga broken bond state. In this case Ga broken bond and $NH_3$ admolecule contribute so they should be treated as donors. The acceptors are $NH_2$ radicals or H adatoms. The electron transfer equation accounts 8 electrons occupying states of both $NH_3$ admolecules and $NH_2$ radicals and 2 at H adatoms, Ga broken bonds are empty:

$$7\frac{3}{4}\beta + 8\frac{3}{4}\alpha + \frac{3}{4}\alpha' + 1\frac{3}{4}\beta' = 8\beta + 8\alpha + 2\beta' \qquad (9)$$

where $\alpha$ and $\alpha'$ are fractions of two types of donors, $NH_3$ admolecules and Ga broken bonds, respectively. The $\beta$ and $\beta'$ are fractions of the two different acceptors, $NH_2$ radicals and hydrogen adatoms, respectively. Naturally the fractions of all types are normalized to unity:

$$\alpha + \alpha' + \beta + \beta' = 1 \tag{10}$$

The above equation maybe simplified to the electron transfer balance,

$$\frac{3}{4}\alpha + \frac{3}{4}\alpha' = \frac{1}{4}\beta + \frac{1}{4}\beta' \tag{11a}$$

or

$$\frac{3}{4}\alpha_{eff} = \frac{1}{4}\beta_{eff}' \tag{11b}$$

where $\alpha_{eff} = \alpha + \alpha'$ denotes summaric donor concentration, reflecting the fact that both donors donate 3/4 electron equally. Similarly acceptor combined concentration $\beta_{eff} = \beta + \beta'$ could be used as both acceptors take 1/4 electron. That simplifies normalization again to:

$$\alpha_{eff} + \beta_{eff} = 1 \tag{12}$$

Thus the ECR state of the surface may be denoted by: $\alpha_{eff} = 0.25$ and $\beta_{eff} = 0.75$. The surface coverage has to be additionally described by fraction of $NH_3$ donors denoted by x and fraction of $NH_2$ acceptors denoted by y. The ECR state is then expressed, as: 0.25 x $NH_3$ ML, 0.75 y $NH_2$ ML and 0.75 (1 - y) H ML. Naturally the fractions of the types of donors (x) and acceptors (y) are within standard interval: $0 \leq x, y \leq 1$.

The second ECR state exists for Fermi level located between Ga broken bond state and CBM. In this case Ga broken bond state becomes acceptor and the only donor is $NH_3$ admolecule. Thus the difference is that Ga broken bond states are occupied by two electrons:

$$7\frac{3}{4}\beta + 8\frac{3}{4}\alpha + \frac{3}{4}\beta'' + 1\frac{3}{4}\beta' = 8\beta + 8\alpha + 2\beta' + 2\beta'' \tag{13}$$

where $\alpha$ is the fractions the donor $NH_3$ admolecules. The $\beta$, $\beta'$ and $\beta''$ are fractions of the three different acceptors: Ga broken bonds, $NH_2$ radicals, and hydrogen adatoms, respectively. Naturally the fractions of all types are normalized to unity:

$$\alpha + \beta + \beta' + \beta'' = 1 \tag{14}$$

The above equation maybe simplified to the electron transfer balance,

$$\frac{3}{4}\alpha = \frac{5}{4}\beta + \frac{1}{4}\beta' + \frac{1}{4}\beta'' \quad (15a)$$

or

$$\frac{3}{4}\alpha = \frac{5}{4}\beta + \frac{1}{4}\beta_{occ} \quad (15)$$

where $\beta_{occ} = \beta' + \beta'$ is the fraction of occupied acceptor sites, $NH_2$ radicals and H adatoms.

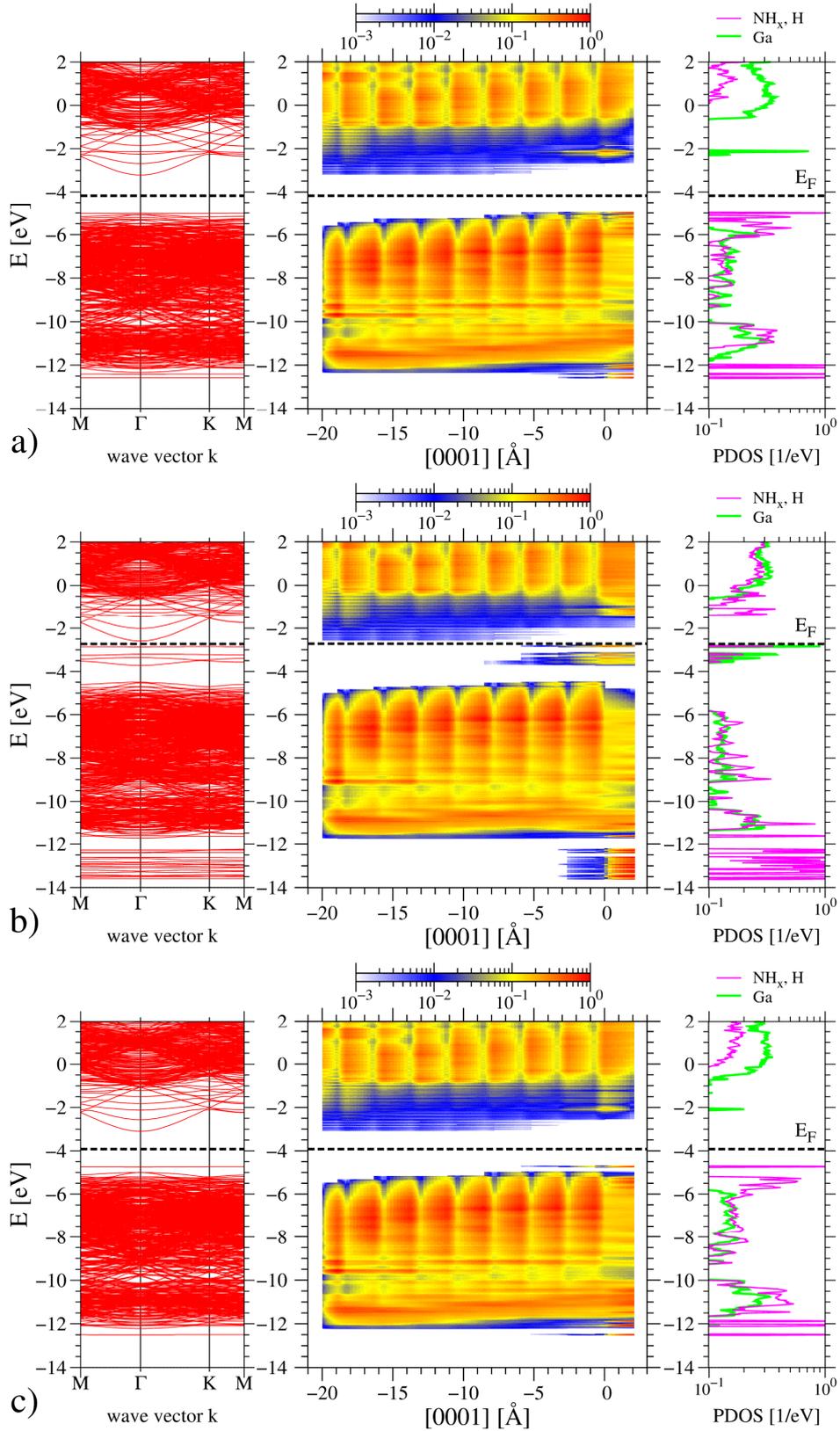

FIG. 9. Dispersion relations (left), space alignment of the bands derived from atom projected density of states (P-DOS) (middle) and DOS projected on surface gallium and nitrogen-hydrogen (in NH$_3$ admolecules) atoms (right) for: (a) 0.125 ML NH$_3$ - 0.25 ML NH$_2$ - 0.5 ML

H; (b) 0.5 ML NH$_3$ - 0.125 ML NH$_2$ - 0.125 ML H (c) 0.125 ML NH$_3$ - 0.375 ML NH$_2$ - 0.375 ML H - covered GaN(0001) surface.

The ECR surface coverage is given by: $\alpha = \frac{1}{4} + \beta, \beta_{occ} = \frac{3}{4} - 2\beta$ parameterized by the fraction of empty sites β, where 0 < β < 3/8. The surface coverage is therefore described as: (0.25 + β) NH$_3$ ML, (0.75 - 2 β) x NH$_2$ and (0.75 - 2 β) (1 - x) H where x is the fraction of NH$_2$ radicals, $0 \leq x, \leq 1$. In Figure 9 the two cases are presented: the upper presenting the first case for 0.125 ML NH$_3$, 0.25 ML NH$_2$, 0.5 ML H and the lower for the second: 0.5 ML NH$_3$, 0.125 ML NH$_2$ and 0.125 ML H.

Dissociative adsorption of ammonia results in creation of NH$_2$ radicals and H adatoms. For small coverage the NH$_2$ radicals may adopt bridge configuration. Thus one possibility is to have partial NH$_2$ bridge coverage. The fraction of NH$_2$ acceptor radicals is denoted by β, thus the number of donors is:

$$\alpha = 1 - 2\beta \tag{16}$$

that accounts the saturation of two broken bonds by the radical in the bridge position. Accounting that the Ga broken bond states are empty and the radical states are occupied by 8 electrons the following electron balance equation is obtained:

$$\beta\left(5 + 2 + 2 \cdot \frac{3}{4}\right) + \frac{3}{4}\alpha = 8\beta + 0\alpha' \tag{17}$$

From these two above equation the ECR NH$_2$ bridge coverage was determined to $\beta = \frac{3}{4}$. Since each bridge configuration covers two sites, the concentration of the radicals fulfills the condition $\beta \leq \frac{1}{2}$. Therefore such ECR configuration could not be realized.

In addition to these configurations, the mixed configurations were verified using ab initio calculation. The following sequence of the configurations recovers both transitions between VBM and Ga- broken and also between Ga-broken bond state and CBM pinned

Fermi level. The first one showing transition between VBM and Ga state is presented in Figure 10.

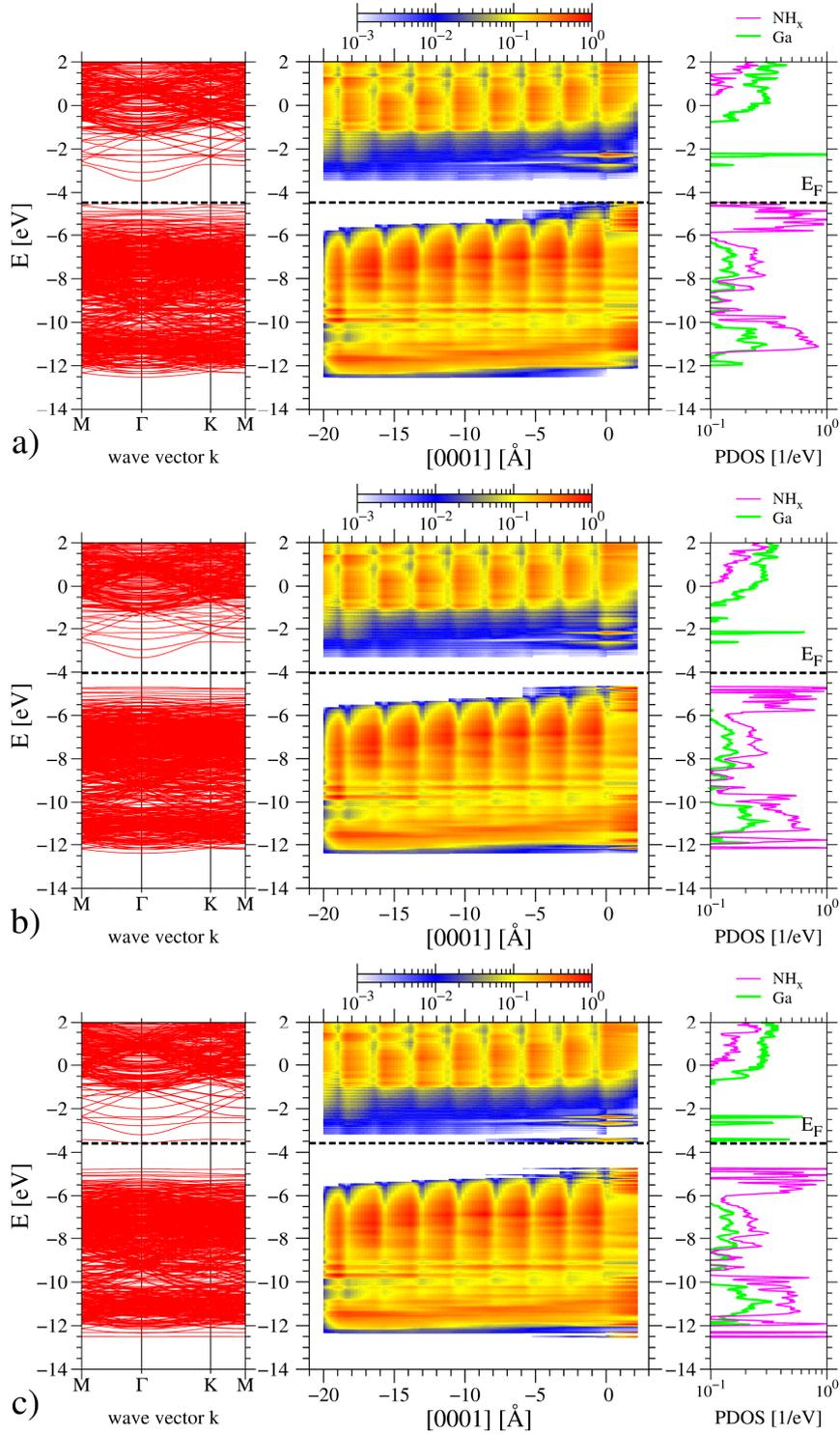

FIG. 10. Dispersion relations (left), space alignment of the bands derived from atom projected density of states (P-DOS) (middle) and DOS projected on surface gallium and nitrogen-hydrogen (in $NH_3$ admolecules) atoms (right), for: (a) 0.8125 ML $NH_2$; (b) 0.0625 ML $NH_3$ - 0.5 ML $NH_2$ (c) 0.125 ML $NH_3$ - 0.6875 ML $NH_2$ covered GaN(0001) surface.

In fact these data once again confirm the transition between these two pinning regimes. The other regime is presented in Figure 11.

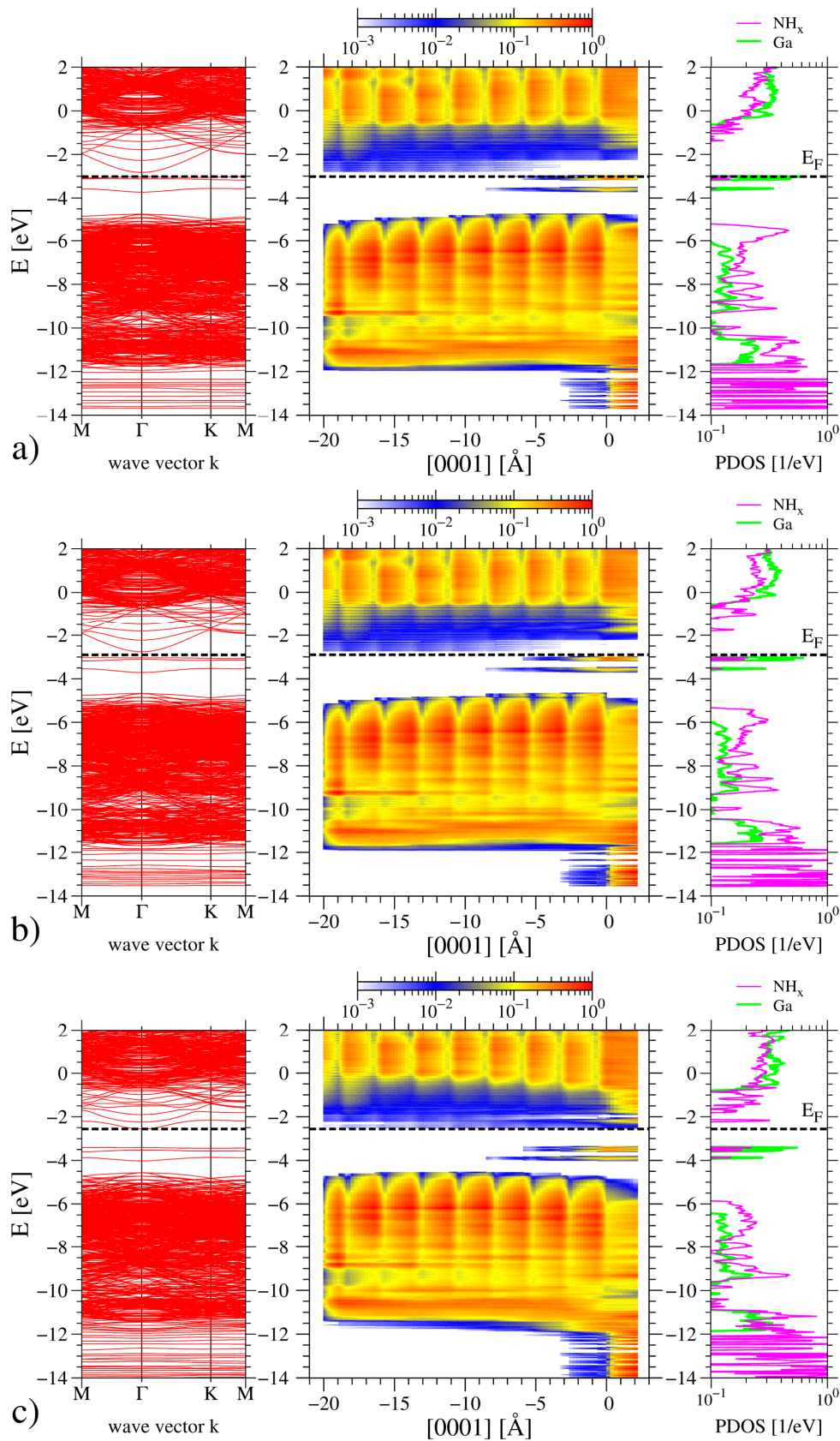

FIG. 11. Dispersion relations (left), space alignment of the bands derived from atom projected density of states (P-DOS) (middle) and DOS projected on surface gallium and nitrogen-hydrogen (in NH$_3$ admolecules) atoms (right) for: (a) 0.375 ML NH$_3$ - 0.4375 ML NH$_2$; (b) 0.4375 ML NH$_3$ - 0.375 ML NH$_2$ (c) 0.5 ML NH$_3$ - 0.3125 ML NH$_2$ - covered GaN(0001) surface.

These data again confirm possibility of transition between these three pinning regimes and validity of ECR rule.

Another possible configuration is that directly related to NH$_3$ dissociation, i.e. coverage by mixture of NH$_2$ radicals and H admolecules. As it was already determined both configurations are characterized by the surface states degenerate with valence band, i.e. both they are acceptors. Denoting the coverage by the radicals by β, and by H adatoms by β', the number of Ga broken bond atoms by α, the following normalization relation is obtained:

$$\alpha = 1 - 2\beta - \beta' \tag{18}$$

Accounting that NH$_2$ radical and H adatom quantum states are occupied by 8 and 2 electrons respectively, the following electron balance equation is obtained:

$$\beta\left(5 + 2 + 2 \cdot \frac{3}{4}\right) + \left(1 + \frac{3}{4}\right)\beta' + \frac{3}{4}\alpha = 8\beta + 2\beta' + 0\alpha \tag{19}$$

From these relations, the total number of acceptor states may be obtained: $\beta + \beta' = \frac{3}{4}$. Naturally these concentrations are limited by the total coverage: $\beta \leq \frac{1}{2}$ and $\beta' \leq 1$. Thus the favorable configuration are these covered by the hydrogen adatoms. As an example the following coverage was tested: $\beta = \frac{1}{8}$ and $\beta' = \frac{5}{4}$. In the (4 x 4) slab that corresponds to 10 H adatoms and 2 NH2 radicals in bridge positions. The remaining two sites are empty. Such configuration is presented in Figure 12.

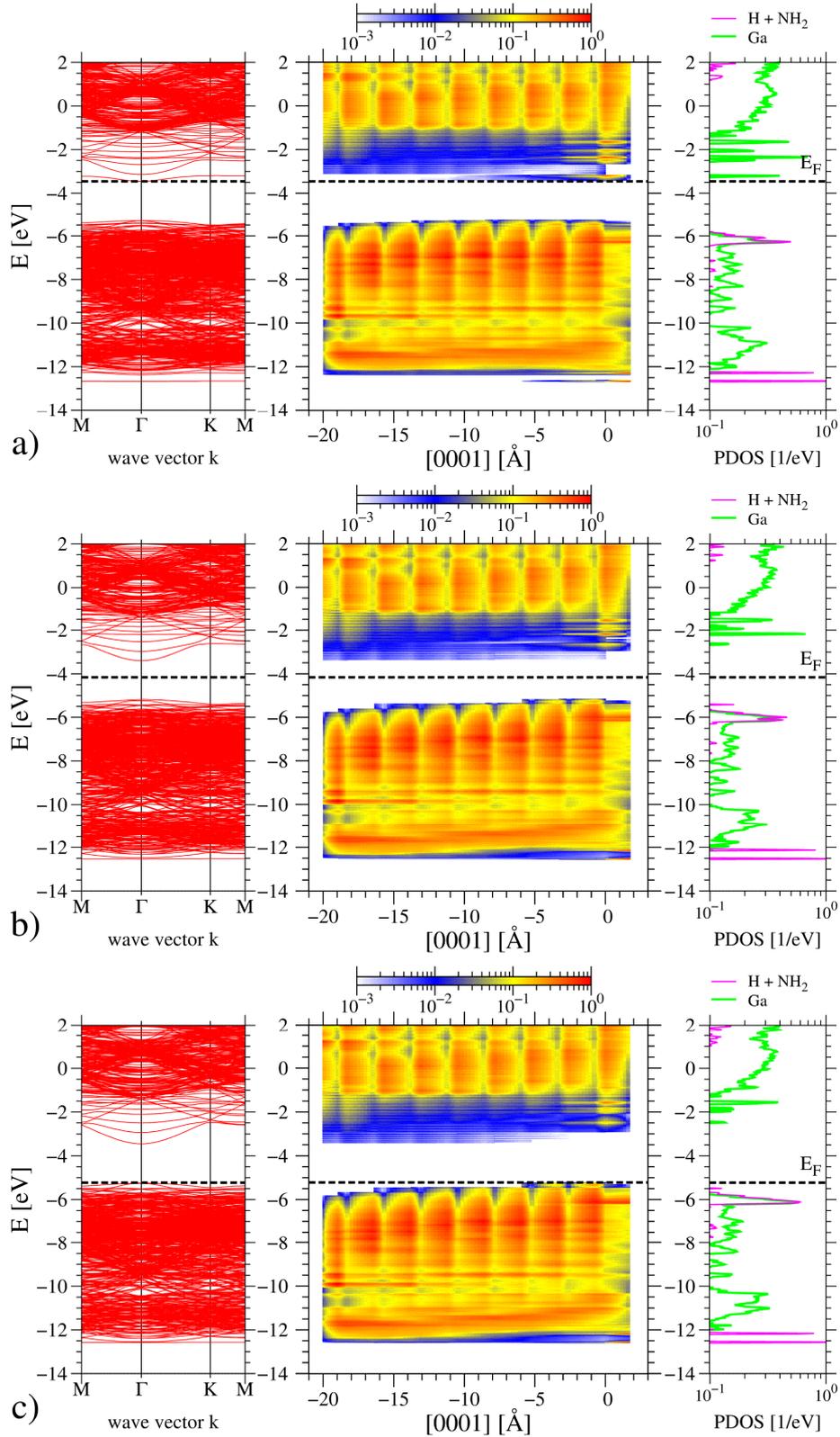

FIG. 12. Dispersion relations (left), space alignment of the bands derived from atom projected density of states (P-DOS) (middle) and DOS projected on surface gallium and nitrogen-hydrogen (in $NH_3$ admolecules) atoms (right) of the N-H and Ga-N bonding (right), for the following coverage of GaN(0001) surface: (a) 0.125 ML NH$_2$ - 0.6875 ML H ; (b) 0.125 ML NH$_2$ - 0.625 ML H (a) 0.125 ML NH$_2$ - 0.5625 ML H.

The results shown in Figure 12 confirm predictions made using ECR rule extending the Fermi level pinning scenario for bridge configurations at GaN(0001) surface. In addition the number of control calculation was made verifying the Fermi level position at the surface for number of ternary coverage, as reported in Table I.

Table I. Fermi level position for ternary NH$_3$ - NH$_2$ - H partial coverage.

| Coverage | Fermi level position |
| --- | --- |
| 0.0625ML NH$_3$ - 0.4375ML NH$_2$ - 0.375ML H | VBM |
| 0.0625ML NH$_3$ - 0.375ML NH$_2$ - 0.4375ML H | VBM |
| 0.125ML NH$_3$ - 0.4375ML NH$_2$ - 0.375ML H | VBM |
| **0.125ML NH$_3$ - 0.375ML NH$_2$ - 0.375ML H** | **Free (VBM/Ga-broken)** |
| 0.25ML NH$_3$ - 0.3125ML NH$_2$ - 0.3125ML H | Ga-broken |
| 0.375ML NH$_3$ - 0.25ML NH$_2$ - 0.25ML H | Ga-broken |
| 0.5ML NH$_3$ - 0.1875ML NH$_2$ - 0.1875ML H | Ga-broken |
| 0.25ML NH$_3$ - 0.25ML NH$_2$ - 0.25ML H | Ga-broken |
| 0.25ML NH$_3$ - 0.1875ML NH$_2$ - 0.1875ML H | Ga-broken |
| **0.5ML NH$_3$ - 0.125ML NH$_2$ - 0.125ML H** | **Free (Ga-broken/CBM)** |
| 0.5625ML NH$_3$ - 0.125ML NH$_2$ - 0.125ML H | CBM |
| 0.625ML NH$_3$ - 0.125ML NH$_2$ - 0.125ML H | CBM |
| 0.6875ML NH$_3$ - 0.125ML NH$_2$ - 0.125ML H | CBM |

Thus the intensive DFT calculations, determining the position of Fermi levels accurately, confirmed the validity of ECR over wide range of surface coverage. The diagram, based on all above calculations, presenting the Fermi level pinning at GaN(0001) surface in ammonia ambient is given in Figure 11. As it is shown the basic three regions exits, for Fermi level pinned at VBM, CBM and Ga broken bond state, with the appropriate borders between regions where the depinning occurs. The ammonia stable configurations, their energies and the adsorption energies will be determined below.

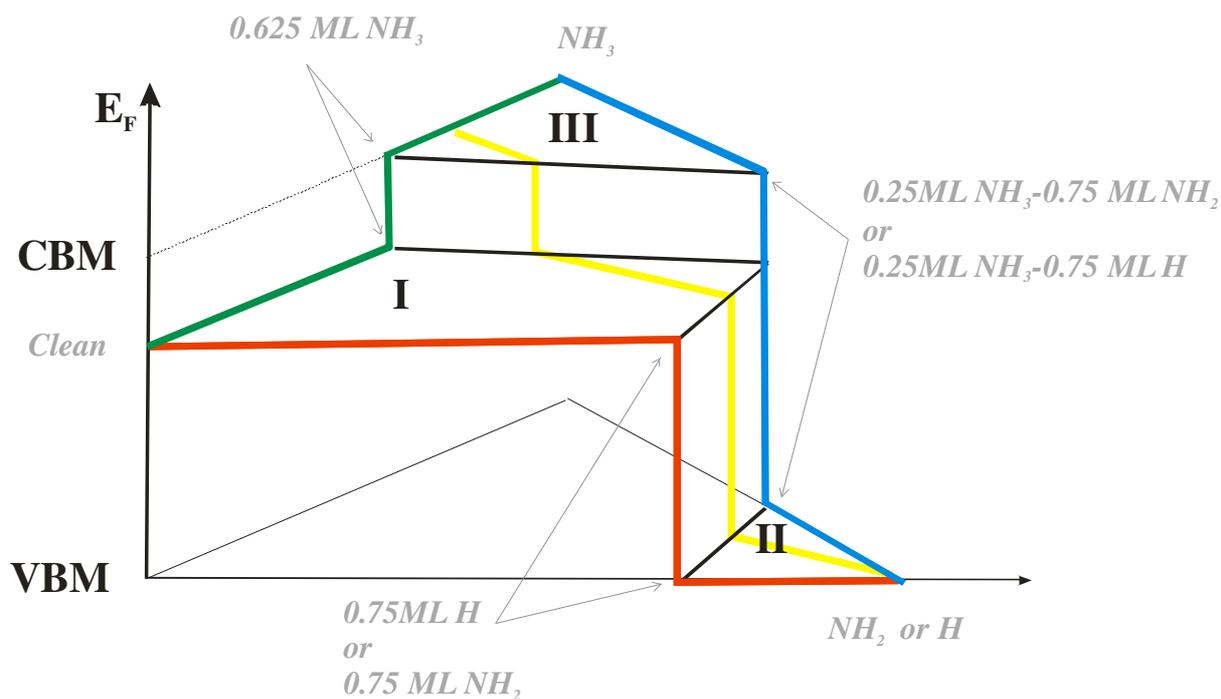

FIG. 13. The nitrogen-hydrogen diagram of Fermi energy pinning at GaN(0001) surface and consequently, the adsorption energy. The three corners represent uniform surface: clean (i.e. absence of any coverage), and the full uniform coverage by H, $NH_2$, or $NH_3$. The broken lines represent the partial unary coverage by H, $NH_2$ or $NH_3$. The full lines represent full binary coverage: H - $NH_2$, H - $NH_3$ or $NH_2$ - $NH_3$. The three regions correspond to following pinning of Fermi level: I - by Ga broken bond state, about 0.6 eV below CBM, II - at VBM, III - at CBM.

### c. $NH_3$ adsorption at empty site

Relatively large (4 x 4) slabs reduces graininess of the simulation of the adsorption processes that attempts to recover the infinite surface. In this study it is assumed that the change of 1/16 of the occupation closely resembles infinitely small variation in the real process. The concentration will be denoted without specific account of the adsorption site which was assumed vacant.

The adsorption of ammonia involving part of the possible configurations, depicted in Figure 13 was already studied recently. The ammonia on the surface partially covered by hydrogen was presented in Refs. 7 and 8. Generally the ammonia adsorption at GaN(0001) surface is barrierless process [8]. The investigations confirmed earlier results indicating that ammonia may be adsorbed dissociatively leading to creation of $NH_2$ radicals located in bridge position and H adatom in the "on-top" position [16-19]. Alternatively as reported also it may be adsorbed molecularly in the "'on-top" position [20]. The results, concerning the adsorption energy may be understood using recently formulated theory of charge transfer at the surface contribution to adsorption energy [6,7]. According to the theory charge transfer may affect the energy even by several electronvolts [6,7]. The electronic charge may be transferred between the bulk states close to the Fermi level which is pinned by surface state [6] or it may be nonpinned [7]. In the first case the adsorption energy depends on the type of the surface pinning state [6] or on the doping in the bulk [7]. In accordance to the diagram presented in Figure 11, the three regions correspond to three types of Fermi level pinning and the transition between them - to Fermi level nonpinned.

According to the results presented in Ref. 7 Fig 5 [7], the adsorption energy of ammonia at GaN(0001) surface partially covered by hydrogen depends on the type of the process. Molecular adsorption energy is close to 2 eV and independent of the hydrogen coverage. This is due to the fact that all 8 states of the molecule adsorbed at the surface are occupied and the charge transfer to these state is impossible. Since these states have their energies deep in the valence band, the opposite transfer is also excluded. The dissociative adsorption leads to creation of 10 states, 8 in $NH_2$ radical and 2 in H adatom. As before both the radical and the adatom states are located in valence band and they could take electrons from the bulk. Accordingly, the adsorption energy is equal to approximately 2eV for low hydrogen coverage, where the Fermi level is pinned at Ga broken bond state (region I in Fig 13) and the energy gain is considerable. For the high hydrogen coverage (region II in Figure 13), the Fermi level is at VBM and the charge transfer contribution to the adsorption energy is small. The energy gain on the adsorption is negative, i.e. the energy has to be added from the crystal to the molecule in order to reach the adsorbed state. The transition between these two region is located in the vicinity of 0.75ML hydrogen coverage, in accordance to Eqs. 2 and 3 and the data presented in Figure 5. The adsorption energy in the ECR region depends on the doping in the bulk [7].

Similar data were obtained for the mixed $NH_2/NH_3$ coverage as presented in Ref 38. It is shown that desorption energy of ammonia depends on the relative $NH_2/NH_3$ ratio [37]. For dominant $NH_2$ content the ammonia is strongly attached to the surface. Up to 0.2 ML of ammonia the desorption energy is equal to 2.0 eV, independent of the $NH_3$ content. For the ammonia content above 0.2 ML the energy falls down to become negative at about 0.4 ML of ammonia. This behavior is in agreement with Eqs 15 and the results presented in Figure 9, where the transition occurs above 0.25 ML of ammonia. The exact concentration depends also on fraction of empty sites.

A third axis in Fig 13 describes the partial unary $NH_3$ coverage, i.e. when the fraction of the sites is covered by ammonia molecules, with the remaining left empty. The energy of ammonia adsorption on such surface is presented in Figure 14.

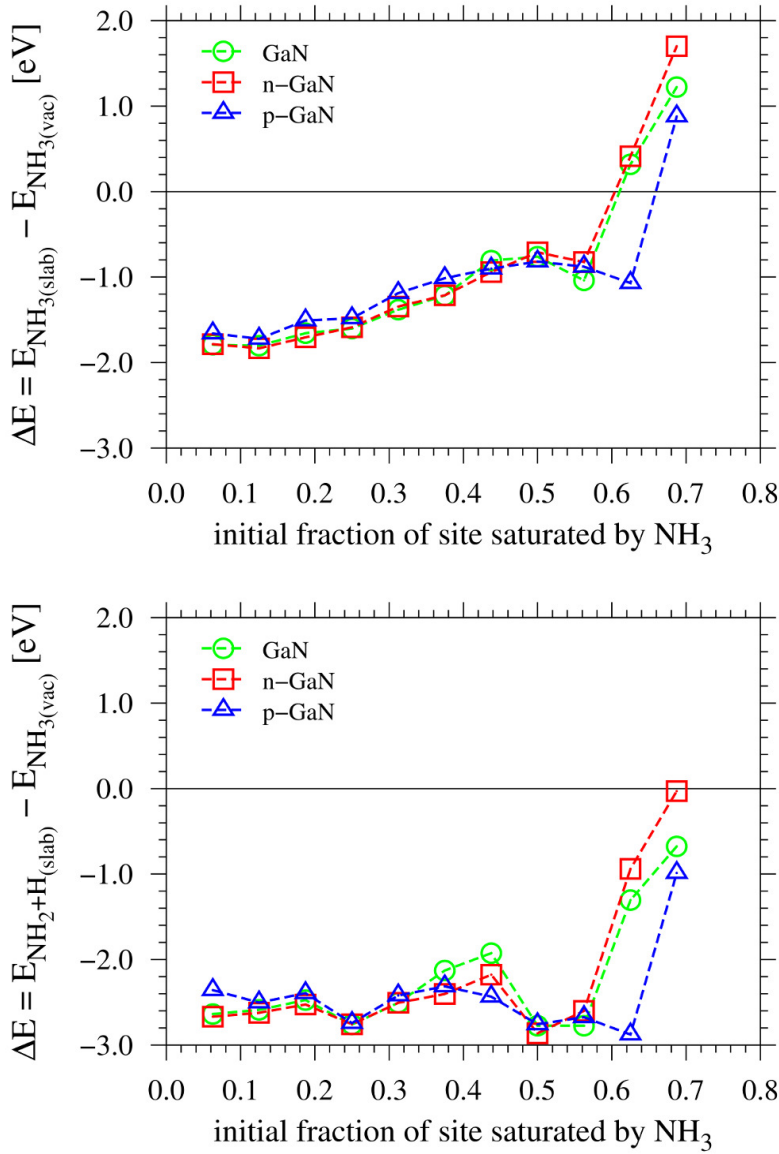

FIG.14. Adsorption energy of ammonia at GaN(0001) surface under partial ammonia coverage: top - molecular, bottom dissociative adsorption, respectively.

Generally, dissociative adsorption has much larger energy, thus the adsorption of ammonia at relatively clean surface leads to dissociation of ammonia into $NH_2$ radical and H adatom. For low coverage, up to 0.3 ML the stable configuration is bridge, for higher - "on-top". The hydrogen adatom is always located in the "on-top" position.

For the coverage above 0.6 ML, the adsorption energy is much reduced for both molecular and dissociative process. Generally molecular configuration is always less stable, thus NH$_2$ coverage is preferred decreasing ammonia content. In fact the increase, similar to the data regarding NH$_2$/NH$_3$ coverage, presented in Ref. 38 could be also explained using the charge transfer theory. In agreement with Eq.8 and data presented in Figure 7, the decrease occurs when Fermi level is shifted from Ga broken bond (region I) to CBM (region III), i.e. at 0.625 ML ammonia coverage. The same applies to data in Ref. 38, and also to Eq. 8 and Figure 8 in this work. In this case the transition from region II to III occurs at 0.25 ML ammonia coverage. The transition reported in Ref. 38 is in good agreement with these data.

The essential point in the charge transfer theory is that the charge transfer may lead not only to increase but also to decrease of the adsorption energy. That happens if the electrons are forced to move to higher energy states because of their excess in the final configuration. Specifically, molecular adsorption of ammonia brings 8 electrons, equal to the number of the new states. Thus all they occupy them. The two Ga broken bond states are occupied by the electrons. Thus these two electrons have to move higher to conduction band which cost the energy. The final energy is increased and the adsorption energy reduced. For the case of molecular these are the two electrons as the ammonia brings eight electrons to 8 new states. For the case of dissociative adsorption, in the case of ammonia molecule located "on-top" these are 8 electrons in NH$_3$ molecule and 4 electrons from the two Ga broken bond states. From these 8 and 2 electrons are located in the NH$_2$ radical and H adatom, respectively. The two electrons are then shifted to conduction bond (CB) states. In the case of NH$_2$ radical in the bridge position, the initial number of electrons is 14. Again 10 are located in the surface states, while 4 other are moved to CB states. Therefore such charge transfer should change the electric state of the surface considerably. The magnitude of the charge transfer may be verified experimentally, using techniques presented in Ref. 39 [38].

The adsorption at mixed configuration is also determined by the Fermi level pinning at the surface. Thus the same energy values are applicable to the regions: I, II and III. From the above determined dependences, given by Eqs. 10-12, the following relation may be obtained for the border between region I and II:

$$\alpha + \alpha' = \frac{1}{4} \qquad (20a)$$

$$\beta + \beta' = \frac{3}{4} \qquad (20a)$$

where α is the coverage by $NH_3$ admolecules, α' = 0.25 - α is the fraction of empty sites. The second relation describes relative occupation by acceptors β and β' are coverages by $NH_2$ radicals and H adatoms. The second border, between region I and III is described by the following relations:

$$\alpha = \frac{1}{4} + \alpha' \qquad (21a)$$

$$\beta = \left(\frac{3}{4} - 2\alpha'\right) x \qquad (21b)$$

$$\beta' = \left(\frac{3}{4} - 2\alpha'\right)(1-x) \qquad (21c)$$

the notation is as above. The relative fraction of $NH_2$ radicals and H admolecules is given by x. The above relations complete description of ammonia at empty sites of GaN(0001) surface.

## C. $NH_3$ adsorption at H-covered site

The adsorption of ammonia at H-covered site of GaN(0001) surface leads to creation of $NH_2$ admolecule and desorption of $H_2$ molecule. The total sequence of reaction could be described as follows.

$$NH_3(g) + H(s) \rightarrow H_2(g) + NH_2(s) \qquad (22)$$

The total energy of such process was determined for the three possible pinning cases: at Ga broken bond, at VBM and CBM. The first case, belonging to region I, was modeled using 4 x 4 slab with the following coverage 0.25ML $NH_3$ - 0.1875ML $NH_2$ - 0.1875ML H. The adsorption energy was -0.158 eV, i.e. it is negative. Thus the conversion is not preferred energetically suggesting that such process is not likely to occur. It has to be noted that leaves 0.375 sites free thus the adsorption is likely to occur at neighboring empty site.

The other to possible cases belong to regions II and III, i.e. Fermi level pinned at VBm and CBM respectively. These regions were modeled using the following configurations: 0.0625ML $NH_3$ - 0.375ML $NH_2$ - 0.375ML H and 0.75ML $NH_3$ - 0.0625ML H. These two regions are characterized by the adsorption energies equal to 0.127 eV and 0.114 eV, respectively. These data indicate that conversion from H to $NH_2$ coverage is effective process in the presence of large amounts of ammonia in the vapor.

## IV. Summary

Adsorption of ammonia at $NH_3/NH_2/H$ covered GaN(0001) surface was analyzed using results of ab initio calculations. The whole configuration space of partially $NH_3/NH_2/H$ covered GaN(0001) surface was divided into the zones of differently pinned Fermi level: at Ga broken bond state for dominantly bare surface (region I), at VBM for $NH_2$ and H covered (region II), and at CBM for $NH_3$ covered surface (region III). It is shown that ECR allows to determine the borders of these regions precisely. The extensive ab intio calculations show validity of ECR for all mixed coverage, necessary to define these three regions.

The adsorption was analyzed using newly identified dependence of the adsorption energy on the charge transfer at the surface. It was shown that ammonia adsorb dissociatively, disintegrating into H adatom and $HN_2$ radical for large fraction of vacant sites while for high coverage the ammonia adsorption is molecular. The dissociative adsorption energy strongly depends on the Fermi level at the surface (pinned) and in the bulk (unpinned) while the molecular adsorption energy is determined by bonding to surface only, in accordance to the recently published theory. The molecular adsorption is determined by the energy of covalent bonding to the surface. Such difference is observed for regions I and II.

It is also shown that ammonia adsorption in region III (Fermi level pinned at CBM) leads to unstable configuration both molecular and dissociative. The drastic change of the adsorption energy is explained by the fact that Ga-broken bond sites are doubly occupied by electrons. The adsorbing ammonia brings 8 electrons to the surface, necessitating transfer of the electrons from Ga-broken bond state to Fermi level. The process is energetically costly so that the total energy is increased, leading to energy increase of both molecular and dissociated processes.

Adsorption of ammonia at H-covered site leads to creation of $NH_2$ radical at the surface and escape of $H_2$ molecule. The process leads to the energy increase in the region I. In this case however the neighboring empty sites are available and ammonia could be adsorbed there. In case of region II and III, the process energy is close to 0.12 eV, thus not large. Nevertheless the inverse process is not possible due to escape of the hydrogen molecule. Thus these processes lead to creation of $NH_2$ coverage.

All these processes are barrierless. Thus in the case of N-rich MOVPE and HVPE processes, the surface is covered predominantly by mixture of $NH_2$ radicals and $NH_3$ admolecules.

## Acknowledgements

The calculations reported in this paper were performed using the computing facilities of the Interdisciplinary Centre for Modelling (ICM) of Warsaw University. This research was supported in part by PL-Grid Infrastructure. The research published in this paper was supported by funds of Poland's National Science Centre allocated by the decision no DEC-2011/01/N/ST3/04382.